\documentclass{article}

\usepackage[final]{style_arxiv}

\setcitestyle{numbers,open={[},close={]}}

\usepackage[utf8]{inputenc} 
\usepackage[T1]{fontenc}  
\usepackage{hyperref}    
\usepackage{url}      
\usepackage{booktabs}    
\usepackage{amsfonts}    
\usepackage{nicefrac}    
\usepackage{microtype}   
\usepackage{graphicx} 		
\usepackage{amsmath}

\title{Delegation to autonomous agents promotes cooperation in collective-risk dilemmas}

\author{%
Elias F.~Domingos \\
  AI Lab\\
  Vrije Universiteit Brussel \\
  1050 Brussels, Belgium \\
  \texttt{eliferna@vub.be} \\
   \And
   Inês Terrucha \\
   IDLab \\
   University of Ghent \\
   B-9052 Ghent, Belgium \\
   \texttt{Ines.Terrucha@UGent.be} \\
   \And
   R\'{e}mi Suchon \\
   ECARES \\
   Université Libre de Bruxelles \\
   1050 Brussels, Belgium \\
   \texttt{Remi.Suchon@ulb.ac.be} \\
   \AND
   Jelena Gruji\'{c} \\
   AI Lab\\
  Vrije Universiteit Brussel\\
  1050 Brussels, Belgium \\
   \texttt{jgrujic@vub.ac.be} \\
   \And
   Juan C. Burguillo \\
   Department of Telematic Engineering \\
   University of Vigo \\
   36310 Vigo, Spain \\
   \texttt{jrial@uvigo.es} \\
   \AND
   Francisco C. Santos \\
   INESC-ID \& Instituto Superior Técnico \\
   Universidade de Lisboa \\
   IST-Taguspark \\
   2744-016 Porto Salvo, Portugal \\
   \texttt{franciscocsantos@tecnico.ulisboa.pt} \\
   \And
   Tom Lenaerts \\
   MLG \\
   Université Libre de Bruxelles \\
   1050 Brussels, Belgium \\
   \texttt{tlenaert@ulb.ac.be} \\
}

\begin{document}

\maketitle

\begin{abstract}
 Home assistant chat-bots, self-driving cars, drones or automated negotiations are some of the several examples of autonomous (artificial) agents that have pervaded our society. These agents enable the automation of multiple tasks, saving time and (human) effort. However, their presence in social settings raises the need for a better understanding of their effect on social interactions and how they may be used to enhance cooperation towards the public good, instead of hindering it. To this end, we present an experimental study of human delegation to autonomous agents and hybrid human-agent interactions centered on a public goods dilemma shaped by a collective risk. Our aim to understand experimentally whether the presence of autonomous agents has a positive or negative impact on social behaviour, fairness and cooperation in such a dilemma. Our results show that cooperation increases when participants delegate their actions to an artificial agent that plays on their behalf. Yet, this positive effect is reduced when humans interact in hybrid human-agent groups. Finally, we show that humans are biased towards agent behaviour, assuming that they will contribute less to the collective effort.
\end{abstract}

\section{Introduction}
\label{sec:intro}
Intelligent autonomous machines/agents are already present at many levels in our daily lives. Examples can be found in a diverse range of domains: autonomous vehicles \citep{yu2016deep,sallab2017deep}, fraud detection \citep{rehak2009dynamic,raj2011analysis,choi2018artificial}, personal home assistants \citep{fruchter2018consumer,kepuska2018next}, call centers \citep{barth2011access}, drones \citep{bartsch2016drones}, recommendation systems \citep{won2014scan} or educational support systems \citep{smutny2020chatbots,fonte2009tq}. These intelligent agents facilitate many of our individual needs while also influencing different aspects of our social lives. Yet, their presence in social contexts raises questions about the role they play in improving or diminishing our willingness to help others or to collaborate in achieving some common future goals. How will human behavior be changed once interactions are mediated by autonomous social agents? Are they used to promote selfish or cooperative actions? Does delegating decisions to agents reduce the sense of individual responsibility? Can agents actually induce cooperation in humans systems, especially when there is a conflict-of-interest between what's good for oneself or good for the group?


In this manuscript, we address some of these questions experimentally, using a game-theoretical perspective. These behavioural economic experiments investigate how human decision-making is affected when people i) delegate their actions to one of a number of pre-defined autonomous agents, provided by third parties and ii) interact in a hybrid human-agent group where participants do not know who is an agent and who is a real person. In the first context, our experiment also examines if having the capacity to "code" or customise the agent with one's own norms affects the observed behavior. While this provides additional insight into the effect of delegation on decision-making, it also provides insight into how people believe the agents should behave when they are acting on their behalf. 

In all experiments --- delegation, programming of the agent, and hybridization --- we adopt the Collective-Risk Dilemma (CRD) \citep{Milinski2008,santos2011risk,Tavoni2011,vasconcelos2014climate,Dannenberg2015,gois2019reward,fernandez2020timing}, a public goods game with uncertain and delayed returns.  The CRD abstracts social decision-making problems such as climate action, abuse of antibiotics or collective hunting, where participants aim to avoid a future loss, rather than to seek a benefit. Notwithstanding these interesting examples from human life, collective risk situations are not exclusive to settings where only humans are involved. There is an increasing number of examples of social cases that involve autonomous agents and that involve different degrees of collective risk. Examples range from self-driving cars to conversational (chat) bots in call centers. 

In the CRD, a group of human participants are confronted with a choice: contribute sufficiently (cooperate) over several rounds to achieve a collective target, ensuring that the group benefit is achieved, or contribute insufficiently (defect) and assume that others will make the contributions to reach the goal, and thus, aim to maximise one’s personal gains. Concretely, in our setup of the CRD, participants, interact in groups of $6$, where each member is required to contribute either $0$, $2$ or $4$ Experimental Monetary Units (EMUs) in every round to a public account out of their private endowment ($40$ EMUs). The experiment consists of 10 rounds and, in the end, if the sum of the total contributions of all players is higher or equal than a collective target of $120$ EMUs, then all players will keep the remainder of their endowment. Otherwise, all players lose their remaining endowment with a probability of $90\%$, which represents the collective risk of the dilemma and thus gain nothing. 


In earlier experiments, \citet{Milinski2008} have evaluated how different risk probabilities affect human decision-making in the CRD, finding that participants only start to contribute when the risk is high ($p=0.9$). Additionally, \citet{fernandez2020timing} repeated the treatment with $90\%$ of risk of \citet{Milinski2008}, adding two additional treatments that investigate how uncertainty about the number of rounds available to reach the collective target affected participants' behaviour. They found that this timing uncertainty increases the polarisation in the population while prompting earlier contributions and reciprocal behaviors in successful groups, which could indicate that under such uncertainty, signaling the willingness to contribute earlier could be key for success. We use the results of their first treatment as control for the experiment presented in this manuscript.

Now, if no humans take part in a CRD, but instead, they select artificial agents to play the game for them, what would be the outcome? To test this, we performed a first treatment in which participants have to delegate their actions during the experiment to an autonomous agent that they select from a small set of possibilities. This experiment examined, on the one hand, which agents are selected and, on the other hand, whether the group of selected agents is more or less successful than humans. In a second treatment, all participants get the same template-agent and are asked to \emph{customise} (rudimentary programming) the agent so it follows an algorithm that represents their beliefs/preferences on how the game should be played. At the same time, we verified what preferences people have concerning these agents, i.e., whether they prefer to play themselves (or not). 

In a third treatment, we examine what happens when the group consists of a combination of humans and agents: is there any difference in the success rate compared to a CRD with only humans or only agents? Again the question depends on the effect of different agent combinations in the group, as different choices are possible. Here we asked what would happen in the “best” case, thus, selecting the agents programmed by participants from successful groups from the previously mentioned customisation experiment (see also details below). Moreover, we evaluate whether participants think agents made more or less contributions than human participants.

The remainder of the paper is organised as follows. In Section \ref{sec:background} gives an overview of the experimental and theoretical literature about delegation and hybrid interactions with autonomous agents. Section \ref{sec:experiemntal_design} introduces our experimental design and in Section \ref{sec:results_discussion} we present and discuss our results. Finally, we summarise our final conclusions in Section \ref{sec:conclusions}.

\section{Background}
\label{sec:background}
The question of how the presence of autonomous agents may affect human social interactions has been explored from sociological, strategical and psychological perspectives \citep{Andras2018,march2019behavioral,camerer2018artificial,DeMelo2018a,Melo2019b,Toll2020,paiva2018engineering,santos2019evolution}. In the context of behavioural economics, where strategic interactions between individuals is analysed, \citet{march2019behavioral} finds, after comparing more than 90 experimental studies, that humans tend to act in a more self-interested or selfish manner when computer players are present, and, often, they are able to exploit such players. Also, \citet{cohn2018honesty} show that individuals tend to cheat about three times more when interacting with agents instead of another human, and that dishonest individuals prefer to interact with agents. Although, as artificial intelligence (AI) evolves, the opposite situation may also occur, i.e., intelligent agents may exploit humans. \citet{camerer2018artificial} argues that behavioural experiments are necessary to understand and predict how this may happen.

Initial results by \citet{Melo2019b} have shown experimentally that delegating to programmable autonomous agents changes the way in which participants approach a social dilemma and is able to promote the evaluation of long-term consequences of choices, which in their case increases cooperation. Also, in another experimental study, \citet{DeMelo2018a} find that participants behave more fairly when acting through autonomous agents in comparison to when human participants directly interact among themselves. They also indicate that when participants program the agents they tend to rely on social norms, which could explain the increase in fairness. 

Introducing agents, programmable/customisable by users to represent their interests in social settings has a number of benefits. The literature commonly refers to these agents as \emph{peer-designed agents} \citep{chalamish2012enhancing,chalamish2012effectiveness,elmalech2014evaluating,manistersky2014development}, and they can be used to reduce costs in the design and testing of mechanisms that involve human decision-makers, such as automated negotiations and mechanism evaluation \citep{lin2010facilitating,elmalech2014evaluating}, agent-based simulations \citep{chalamish2012effectiveness} or testing experimental hypothesis in game-theoretic scenarios. At the same time, those applications assume that the peer-designed agents truly represent their designer strategy/behaviour. However, this may not always be true and may depend on the context of the decision-making scenario. 

For instance, \citet{grosz2004influence}, \citet{elmalech2014can} and \citet{manistersky2014development} find that, in a negotiation (or trading) setting, participants design agents closer to rational game-theoretic models, instead of more fair or cooperative agents, which contrast with human behaviour observed in direct interactions \citep{straub1995experimental,andreoni2006testing} and the increased fairness observed by \citet{Melo2019b}. Also, other authors find that programmed agents can increase the ability of participants to coordinate when negotiating \citep{raz2009investigating}, which can be an useful trait for our case study in the CRD.

When it comes to hybrid interactions between humans and autonomous agents, as previously mentioned, the game-theoretic literature indicates that humans tend to act more selfishly in those scenarios \citep{march2019behavioral}. However, \citet{DeMelo2019} show experimentally that, by adding visual emotional and cultural cues to the agents, this cooperation barrier against the agents can be surpassed. Concretely, the authors find that while cultural cues cause similar bias as those observed against humans, when agents express emotional cues, participants cooperated with them just as much as against other humans. This research also hints that trust is a very important factor in human-agent interactions. In this line, \citet{Andras2018} suggests that a framework of trust between humans and autonomous agents must be developed. Moreover, it also prompts further research on how such trust currently affects hybrid strategic interactions. \citet{han2020trust} argues that trust-based strategies can be very effective in human-agent interactions, particularly when there is a lack of transparency about the agents' strategies.

In general, experimental results appear to be dependent on the context of the strategic interaction, generating two opposing hypotheses: i) interaction with or through autonomous agents affects social preferences in such a way that it promotes more selfish behaviours and thus decreases cooperation; ii) interacting through and with autonomous agents may eliminate or reduce negative effects of emotions such as revenge or \emph{fear of betrayal} \citep{Dana2006a,Declerck2013,Brock2013,Exley2016a,Bellucci2016}, favoring pro-social behaviours and thus the increase cooperation.

\section{Experimental design}
\label{sec:experiemntal_design}

To explore experimentally the questions exposed in the introduction, we designed three experimental treatments that test how the presence of autonomous agents influence participant behaviour. These treatments are compared to a control experiment in which only humans participate. 

In all treatments, participants play a CRD in a group of $6$ members for 10 rounds with a collective target of $T=120$ EMUs. Each participant starts with a private endowment of $E=40$ EMUs and will lose their remaining endowment at the end of the game with probability $p=0.9$ if the joint contributions of the group do not reach the collective target.

\subsection{Control treatment with only humans}
\label{sec:Hexp}
In the control experiment, human participants from each session are randomly sampled in groups of $6$. All their decisions are anonymous, i.e., neither the other participants nor the experimenters know the identity of the participant making the decisions. The experiment is held in a laboratory and participants interact with each other through a computer interface. In each round, each participant is able to observe their own contribution in the previous round and the contributions of the other participants, thus, allowing them to adopt conditional decisions. At the end of the experiment, participants are shown the result of the experiment. In case their group did not reach the target, they are shown the exact result as a function of the $p$, i.e. a randomly generated value determines whether they lose or win their remaining endowment. Finally, they have to complete a short survey that elicits their motives during the experiment. The full description of the results of this treatment are presented in \citep{fernandez2020timing}.

\subsection{\emph{Delegate} treatment (T1)}
\label{sec:delegate}
\begin{figure}[htbp]
 \centering
 \includegraphics[width=0.99\linewidth]{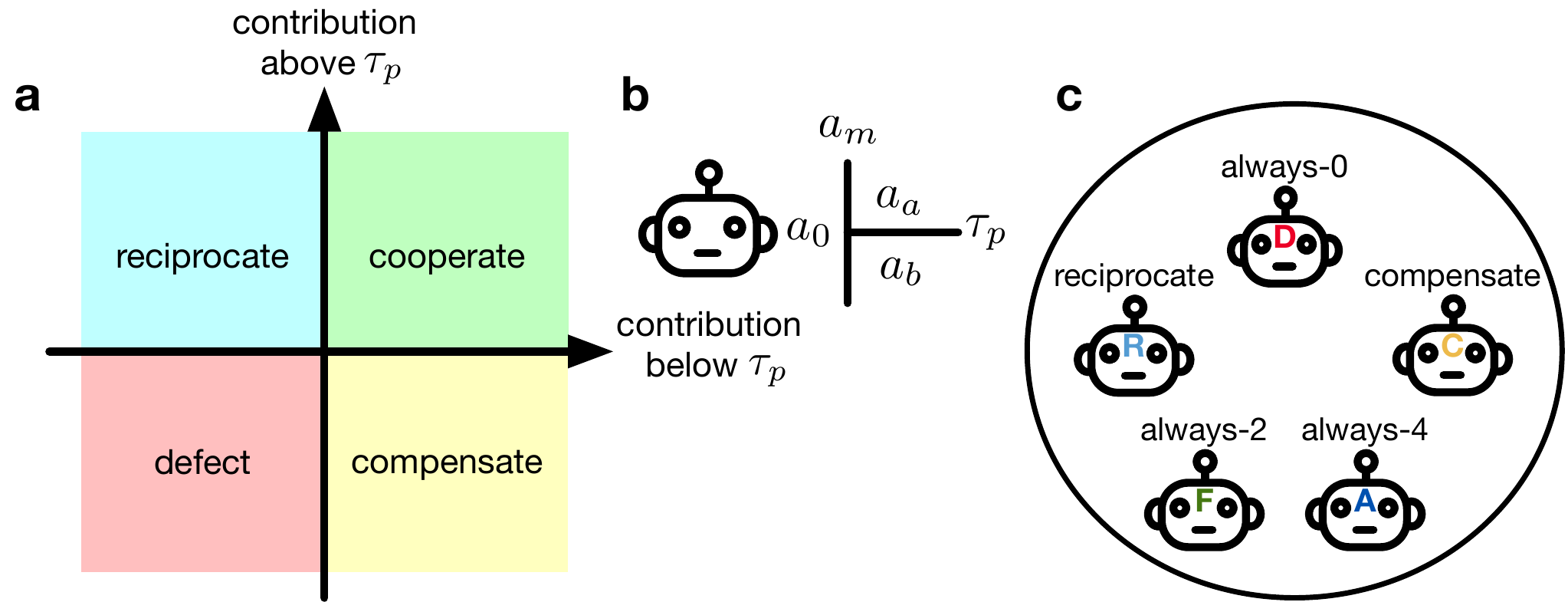}
 \caption{Representation of agent behaviour. Panel a) classifies the behaviour of agents in relation to how much they contribute in response to a stimuli that is above or below a certain personal threshold $\tau_p$. When an agent contributes the same positive amount for stimuli below or above the threshold, we say that it cooperates. When it contributes more for values below (above) the threshold, we say that the agent compensates (reciprocates). Finally, when the agent does not contribute, we say that it defects. Panel b) shows a general structure for the description of the agent's behaviour. An agent has 5 parameters that consist on an initial contribution ($a_{0}$), a personal threshold for the external stimuli ($\tau_{p}$), and the contributions to be made when the external stimuli is above ($a_{a}$), equal ($a_{m}$), or below ($a_{b}$) $\tau_{p}$. Panel c) represents the pool of agents that can be selected in the \emph{delegate} treatment. Participants may select between cooperative agents (e.g., always-2, always-4), conditional agents (e.g., reciprocal, compensator) and defecting agents (e.g., always-0), however, these names are not shared with them, and agents are named with letters from the alphabet (e.g., A, B, etc.).}
 \label{fig:agent_types}
\end{figure}

Before the experiment starts, participants are requested to choose one of 5 types of artificial agents to play the game in their place (see Figure~\ref{fig:agent_types}c). Thus, in this treatment, participant make a single decision in the experiment. For convenience, we attribute names to the available agent behaviours, i.e., always-0, always-2, always-4, compensator and reciprocal. However, during the experiment, agents were only presented by an alphabet letter (e.g., A, B, etc.) and there was never any framing related to the fairness of the agent’s behaviours. Afterwards, they are able to observe the decisions of their agent and the other agents in the group throughout the game. They can also observe the content of the public and private accounts. At each round, participants have a maximum of 10 seconds to observe the information in the screen. Once the game ends, the result is shown on the screen, including information about the participant's final payoff. Finally, participants are requested to complete a short questionnaire that helps elucidate their decision to select a particular agent and how they experienced the performance of their agent.

\subsubsection{Description of the agents}

The first three agents (\emph{always-0}, \emph{always-2}, \emph{always-4}) display unconditional behaviours, i.e., they contribute in every round 0, 2, or 4 EMUs respectively. The last two exhibit conditional behaviours, which respond to external stimuli, in this case, the total contributions of the group members in the previous round ($a_{-i}(t-1)$), in function of a personal threshold $\tau_{p}$, which defines a trigger for behavioural change in the autonomous agents. Agents compare external stimuli to this threshold and adapt their contribution levels depending on the comparison result.

In this treatment, we assume $\tau_{p}=10$ EMUs, representing the total amount of contributions in one round made by the other participants. We call this a locally fair amount since if participants contribute this amount in every round and you also contribute $2$ in every round then the threshold of $120$ EMUs would be reached by the end of the game and everyone will have the same gains to take home. Given this threshold $\tau_p$, \emph{Compensators} only contribute if $a_{-i}(t-1) \leq \tau_{p}$. In contrast, \emph{reciprocal} agents only contribute if $a_{-i}(t-1) \geq \tau_{p}$. All autonomous agents stop donating once the collected contributions reach (or surpass) the collective target. These agent types emulate the behaviors identified in the \emph{human} control treatment \citep{fernandez2020timing}.

\subsection{\textit{Customise} treatment (T2)}
\label{sec:customise}
In the customization treatment, participants are requested to configure an artificial agent that will act in their place during the experiment. Again, in this treatment, participants make a single decision in the experiment. Each artificial agent can be customised in such a way that it can adapt its action to the actions of the other agents within the group, i.e., participants make a conditional agent. To do this, participants are requested to configure the values of 5 parameters (see Figure~\ref{fig:agent_types}b) that define an agent’s behaviour:

\begin{itemize}
 \item Threshold ($\tau_{p} \in[0..20] \text{ in steps of 2}$): is an integer that will be compared to the total contribution of the other group members in the previous round, i.e., the agent will select one of the three following actions based on whether the contributions of the other members of the group in the previous round ($a_{-i}(t-1)$) are greater, equal or lesser than.
  \item Initial action ($a_{0}\in{0,2,4}$): the action the agent will take in the first round of the game.
  \item Action above ($a_{a}\in{0,2,4}$): the action the agent will take if $a_{-i}(t-1)>\tau_{p}$.
  \item Action equal ($a_{m}\in{0,2,4}$): the action the agent will take if $a_{-i}(t-1)=\tau_{p}$.
  \item Action below ($a_{b}\in{0,2,4}$)): the action the agent will take if $a_{-i}(t-1)<\tau_{p}$.
\end{itemize}

Allowing participants to configure their agent, effectively increases the behavioural space that can be displayed in the CRD in comparison to the delegate treatment. Concretely, there is a maximum of $10*3^{4}=810$ possible behavioural combinations in which an agent can be configured. Thus, if we observe an increase in collective success in both treatments, we can rule out that this result is caused by a limited number of choices, and instead we may conclude that it is an effect of delegating to autonomous agents. Finally, in order to assess the level of trust in the delegation system, we ask participants once the experiment has finished, whether, if given the option in another experiment, they would decide to delegate/customise again, or play the game themselves.

\subsection{\textit{Nudge} treatment (T3)}

In the hybrid experiment, we performed the same setup as described in the human control treatment (see Section \ref{sec:Hexp}) . Except that in this case, each group is formed by 3 human participants and 3 artificial agents. Participants are told that the agents have been designed to adopt a human-like behaviour in the CRD experiment: we selected the agents’ behaviours from the pool of agents that were part of successful groups in the customise treatment (see Section \ref{sec:customise})). Since we select only agents programmed by participants from successful groups (of the customise treatment), our hypothesis is that those agents may be able to \emph{nudge} human participants into cooperative behaviors. 

At the end of the experiment, we test whether participants were able to discern which members of the group were human and which were not. The participants are not informed who the agents are in the session in order to ensure that they are just responding to the actions and not the fact that the action is performed by a human or an agent.

\subsection{Sample and collected data}

A total of $246$ participants were part of this experiment, of which $174$ participated in the 3 treatments that involve autonomous agents. The experiment was carried out at the \emph{<removed for reasons of anonymity>}. The population was mainly composed of university students between the ages of $19$ and $33$, of which $49.4\%$ were female. The \emph{delegate} treatment had $n=90$ participants divided into $15$ groups of $6$ participants each. The \emph{customise} treatment had $n=48$ participants and $8$ groups. Finally, the \textit{nudge} treatment had a total of $n=48$ participants and 16 hybrid groups. Moreover, the \emph{human} control treatment was part of another experiment presented in \citep{fernandez2020timing} and consisted of $n=72$ participants divided into $12$ groups of $6$ participants. Due to the COVID-19 situation, we were unable to perform more sessions of the \textit{customise} and \textit{nudge}. Nevertheless, our current data already hints towards interesting phenomena.

For all treatments, we stored the result of the experiment, the configuration over all rounds and the response to a final questionnaire. In \textit{delegate} (and \textit{customise}) treatment we also stored, for each participant, the selected (customised) agent and the time taken to select it. For the \textit{nudge} treatment, we collected the contribution at each round, the actions of the other members of the group in the previous round, the content of the private and public accounts at each round, the predictions of the total content of the public account after each round, and the time taken to make each decision. We also stored the responses to a follow-up survey that participants had to complete once the experiment was finished. In this regard it is important to mention that $1$ participant has been excluded from the survey results in the delegate treatment, since the participant did not complete the survey.

\begin{figure}[htbp]
 \centering
 \includegraphics[width=0.5\linewidth]{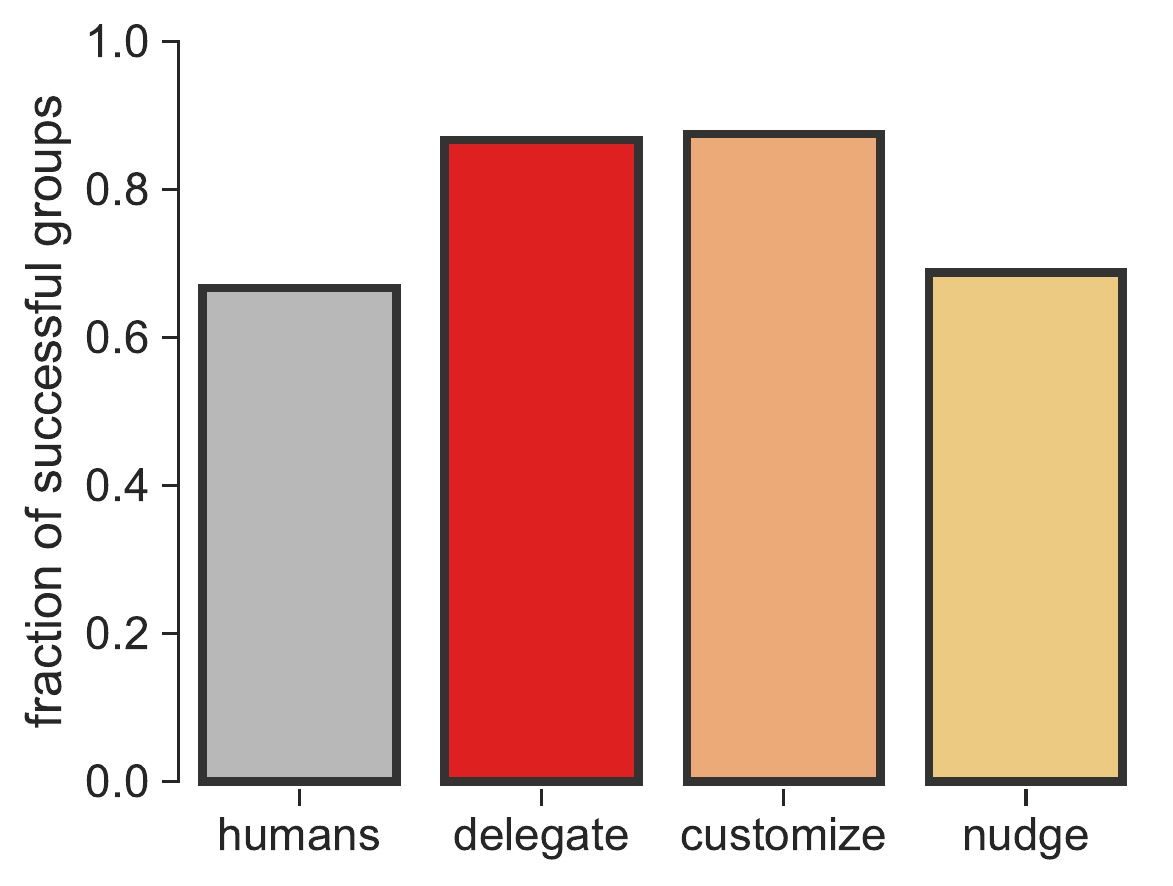}
 \caption{Group achievement and average payoff. The group achievement obtained in the \emph{humans} control treatment (black: $n=12$) is compared to the results of the \emph{delegate} (red: $n=15$), \emph{customise} (orange: $n=8$) and \emph{nudge} (yellow: $n=16$) treatments. There is an increase in groups that achieve the target on the \emph{delegate} and \emph{customise} treatments, however this trend is not maintained in the \emph{nudge} treatment.}
 \label{fig:group_achievement}
\end{figure}

\section{Results and Discussion}
\label{sec:results_discussion}

The fraction of successful groups (group achievement, $\eta$) is an important indicator of the cooperative response in the CRD. In the treatment with only \emph{humans}, $66.7\%$ of the groups successfully achieved the collective target, despite the high level of risk ($p=0.9$). In a previous CRD experiment, \citeauthor{Milinski2008} indicate an even lower result ($\eta=0.5$). In contrast, Figure~\ref{fig:group_achievement} shows that the fraction of successful groups increases in the \emph{delegate} and \emph{customise} treatments ($87\%$ and $87.5\%$ respectively), which indicates that delegation has a positive effect on group success. However, this trend does not persist in the \emph{nudge} treatment, where participants interact with agents and not through them.

\begin{figure}[ht]
 \centering
 \includegraphics[width=0.99\linewidth]{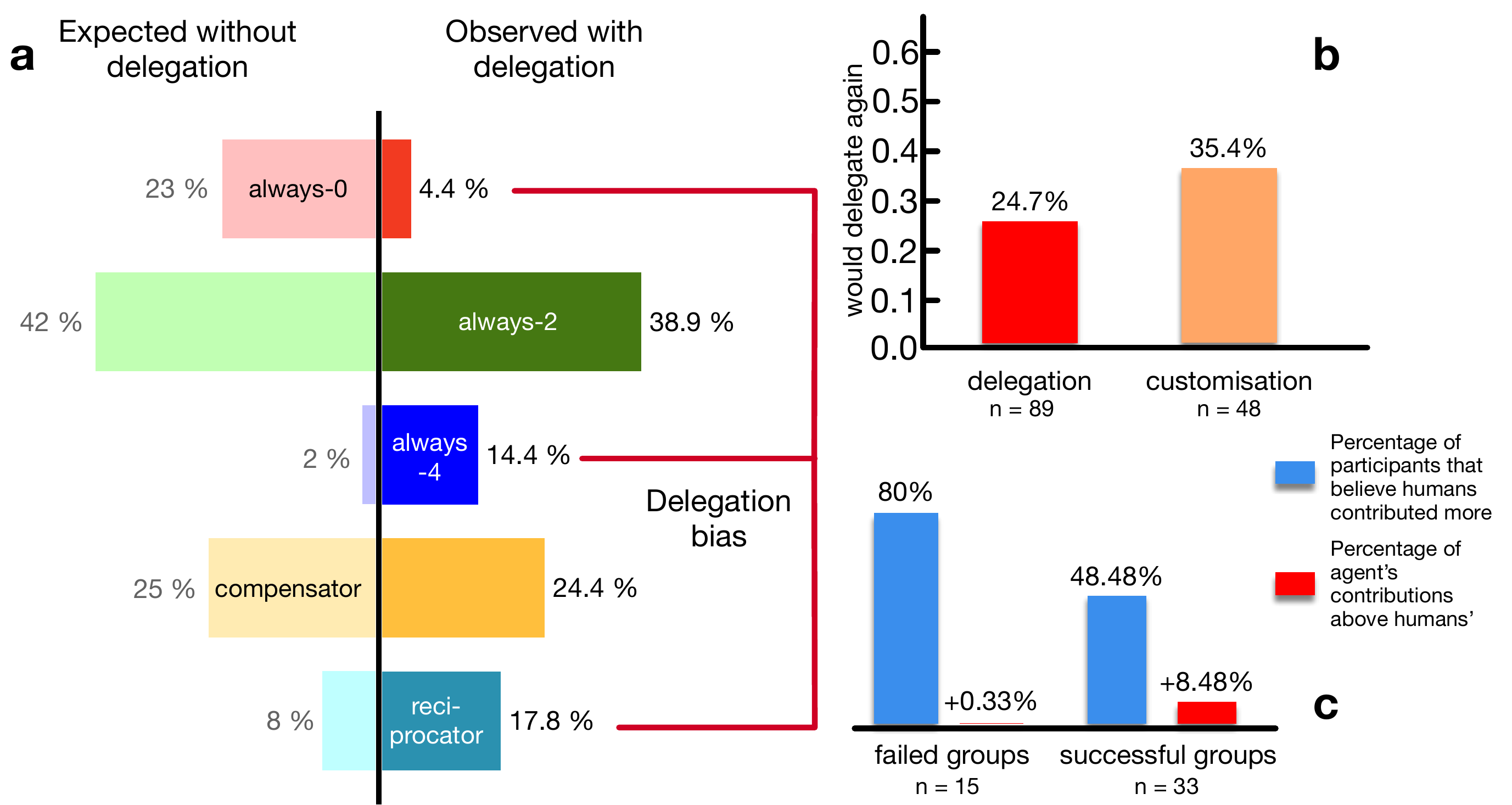}
 \caption{Agent choice bias and participant response. Panel (a) shows the distribution of agent behaviors chosen by the participants (right of the black line) in the \emph{delegate} treatment ($n=89$) in comparison to the expected values predicted by the model described in \citep{fernandez2020timing} (left of the black line). The observed values indicate that there is a bias toward behaviors that are beneficial for the success of the group. Panel (b) shows the percentage of participants that would choose to delegate/customize an agent to play in their place if they were given the option to participate again in the experiment by themselves or to choose an agent. These results show that most participants would rather participate in the experiment by themselves, despite the higher rates of group success in comparison to the experiments with only humans. Nevertheless, there is an increase in participants that would delegate again in the \emph{customise} experiment ($n=48$). This could indicate that allowing the customisation of artificial agents in CRD systems could be a potential relevant design feature to increase their penetration across users. Panel (c) shows the percentage of participants who consider that humans had made the greatest contribution effort during the \emph{nudge} treatment, compared to the average contribution difference of agents with respect to humans in a group. These results show that while $80\%$ of the human players from failed groups ($n=15$) believed humans contributed more during the experiment, artificial agents had marginally contributed more ($+0.33\%$). Such a result indicates a strong perception bias that attaches guilt of failure to the artificial agents. This perception decreases in participants from successful groups, where only $48.48\%$ believe humans had put most of the effort ($n=33$), yet the extra effort made by the agents is only and $8.48\%$ higher than that of the human players.}
 \label{fig:agent_bias}
\end{figure}

Also, if we look at the distribution of selected agent behaviours (see Figure~\ref{fig:agent_bias}a, right of the black line) in the delegate treatment, we can observe that always-2 and compensator were the behaviours selected more frequently. This is in agreement to the predictions of the model of the experimental results of the \emph{humans} control treatment (left of the black line) described in \citep{fernandez2020timing}. Yet, the predicted and observed results for the \emph{always-0}, \emph{always-4} and \emph{reciprocal} agents differ, with an increased preference for cooperative strategies (\emph{always-4} and \emph{reciprocal}) and a decrease in the fraction of \emph{always-0}. We attribute this difference to a delegation bias towards behaviours that are beneficial to the collective, which is corroborated by the answers on the surveys (most participants explain that their motivation was to maximise the probability of group success as opposed to their personal benefit).

The literature on behavioural economics \citep{Dana2006a,Declerck2013,Brock2013,Exley2016a,Bellucci2016} theorises that when interacting in a group of humans, participants tend to be afraid of being betrayed or cheated by the other participants. This fear of betrayal, or \emph{betrayal aversion}, is reduced when humans act through autonomous agents \citep{DeMelo2017,DeMelo2018a,Melo2019b}. Additionally, the chosen/customised agent cannot be changed during the experiment. Therefore, participants are forced to think about the future consequences of their choice, transforming delayed rewards into immediate ones. These several factors play an important role in the observed increase of cooperation.

Moreover, at the end of the delegation experiments, participants were asked whether, if given the option to play again, they would choose to delegate once more to an artificial agent, or to play the game themselves. Figure~\ref{fig:agent_bias}b shows that the majority of participants in the \emph{delegate} and \emph{customise} treatments would prefer to play the game themselves, even though the results from the \emph{humans} treatment indicate that this would lead to lower levels of collective success. This choice suggest a reticence to give away agency, and a preference for retaining control over the actions at each round. Yet, in the \emph{customise} treatment, there was an increment in the fraction of participants that would delegate, suggesting that the ability to customise/configure the agent, increases participants’ trust in it, which is a relevant feature when designing delegation systems. Therefore, future research should explore how delegation and customization should be combined to achieve both cooperation and user satisfaction.

Finally, regarding the \emph{nudge} treatment in which humans interact in hybrid groups with autonomous agents, our current data shows that there is a significant proportion of participants that believe agents make less effort than human participants, even though this is not true. Concretely, in Figure~\ref{fig:agent_bias}c, we show that the majority of participants from failed groups ($80\%$, $n=15$) consider that humans have made more contributions than agents, when in reality artificial agents have marginally contributed more ($+0.33\%$). At the same time, in successful groups, $48.48\%$ of participants ($n=33$) still believe that humans are making more effort than agents, while in this case agents have made $8.48\%$ more contributions than humans. This implies that there exists a negative perception towards artificial agents, i.e. participants associate a higher degree of selfishness to them than to humans. On the other hand, and however we would require another experiment to prove this hypothesis, a hybrid group seems to focus human attention on the agents, indirectly producing a similar reduction in betrayal aversion to what has been observed in the delegation treatments. 

\section{Conclusions}
\label{sec:conclusions}

Previous research has shown that people, when delegating to other humans, tend to choose those who act more selfishly \citep{Exley2016a, Bellucci2016}. In contrast, observations throughout this experiment suggested that people prefer to delegate their decisions to agents that benefit the collective, thus increasing cooperation. This has also been observed in previous experiments with autonomous agents \citep{Melo2019b}. Yet, subjects still prefer to retain control of their actions. However, being able to customise their agent, increased the fraction of participants that were willing to delegate. But, when interacting with these customised agents, subjects in hybrid groups tended to assign the responsibility of failure to the autonomous agents. Therefore, further investigation should address the interplay of social expectations towards autonomous agents, when delegating and in hybrid interactions, with the ability to embed preferences into the agents through customisation.



\section*{Broader Impact}

The experimental study presented in this manuscript offers insights into the impact of delegation towards autonomous agents and hybrid collective systems. Our work shows that the use of artificial delegates, in cases where human fears like betrayal aversion may negatively influence the outcome, can improve the chances of a positive outcome. Defining the decision-making process for future gains upfront and explicitly makes participants think about what they want, as opposed to emotionally responding to cues coming from other participants in the games. Yet, still, it is not enough to reach a $100\%$ success. Communicating what agent a participant selects or informing others about the customisation of a participant's agent, may ensure that this last hurdle is overcome, just as in the use of pledges \citep{Tavoni2011,Dannenberg2015}. The similarity of the CRD with many real-world situations also ensures that the conclusions we draw from these experiments have a wide impact, hopefully helping to achieve fundamental understandings.

The current results have natural limitations associated with the the demographics of the participants pool as well as its size. As it is often the case in this type of studies, it would be meaningful to explore how other age-groups would deal with delegation to autonomous systems and to expand further the number of sessions for the the nudge and customisation treatments. We have indications that children respond differently. Such expansions will allow us to remove biases in the experimental results, helping us to find further support for the observations we have made. Moreover, although our experiments hint towards a relationship between trust and customisation, a different experiment is required to provide a deeper understanding of the mechanisms behind this result. Finally, our study has focused on a scenario of high collective risk. We believe it would be important to observe whether the levels of cooperation and human biases remain unchanged in scenarios of intermediary or lower level of risk, or whether participants' preferences towards intelligent agents are risk dependent.

\begin{ack}
E.F.D. is supported by an F.W.O. (Fonds Wetenschappelijk Onderzoek) Strategic Basic Research (SB) PhD grant (nr. G.1S639.17N), I.T. is supported by the DECISIOs project funded by the F.W.O. grand nr. G054919N. R.S. is supported by an FNRS grant (nr. 31257234). J.G. is supported by an F.W.O. postdoctoral grant. T.L. is supported by the F.W.O. project with grant nr. G.0391.13N, the F.N.R.S. project with grant number 31257234 and the FuturICT2.0
(www.futurict2.eu) project funded by the FLAG-ERA JCT 2016. J.G. and T.L. are supported by the Flemish Government through the AI Research Program. F.C.S. acknowledges support by FCT-Portugal (grants PTDC/CCI-INF/7366/2020, PTDC/MAT/STA/3358/2014, and UIDB/50021/2020). J.C.B is supported by Xunta de Galicia (Centro singular de investigación de Galicia accreditation 2019-2022) and the European Union (European Regional Development Fund - ERDF).
\end{ack}

\end{document}